\documentclass{iau} 
\usepackage{graphicx}

\title[IAUS316.~~Binaries in young massive star clusters] 
{The dynamical importance of binary systems in young massive star clusters}

\author[Richard de Grijs, Chengyuan Li \& Aaron M. Geller]   
{Richard de Grijs$^1$, Chengyuan Li$^{1,2}$, \and Aaron
  M. Geller$^{3,4}$}

\affiliation{$^1$Kavli Institute for Astronomy \& Astrophysics and
  Department of Astronomy, Peking University, Beijing, China \\ email:
  {\tt grijs@pku.edu.cn} \\[\affilskip]
$^2$Purple Mountain Observatory, Chinese Academy of Sciences,
  Nanjing, China \\[\affilskip]
$^3$Center for Interdisciplinary Exploration and Research in
  Astrophysics (CIERA) and Department of Physics and Astronomy,
  Northwestern University, Evanston, IL, USA \\[\affilskip]
$^4$ Department of Astronomy and Astrophysics, University of Chicago,
  Chicago, IL, USA}

\pubyear{2015}
\volume{316}  
\jname{Formation, Evolution, and Survival of Massive Star Clusters}
\editors{C. Charbonnel \& A. Nota, eds.}
\begin{document}

\maketitle

\begin{abstract}
Characterization of the binary fractions in star clusters is of
fundamental importance for many fields in astrophysics. Observations
indicate that the majority of stars are found in binary systems, while
most stars with masses greater than $0.5 M_\odot$ are formed in star
clusters. In addition, since binaries are on average more massive than
single stars, in resolved star clusters these systems are thought to
be good tracers of (dynamical) mass segregation. Over time, dynamical
evolution through two-body relaxation will cause the most massive
objects to migrate to the cluster center, while the relatively
lower-mass objects remain in or migrate to orbits at greater
radii. This process will globally dominate a cluster's stellar
distribution. However, close encounters involving binary systems may
disrupt `soft' binaries. This process will occur more frequently in a
cluster's central, dense region than in its periphery, which may mask
the effects of mass segregation. Using high resolution {\sl Hubble
  Space Telescope} observations, combined with sophisticated $N$-body
simulations, we investigate the radial distributions of the
main-sequence binary fractions in massive young Large Magellanic Cloud
star clusters. We show that binary disruption may play an important
role on very short timescales, depending on the environmental
conditions in the cluster cores. This may lead to radial binary
fractions that initially decline in the cluster centers, which is
contrary to the effects expected from dynamical mass segregation.
\end{abstract}

\firstsection 
\section{Introduction}

Star cluster physics has a fundamental bearing on many important
fields in contemporary astrophysics. Applications range from their use
as tracers of the most violent episodes shaping galaxies and their
evolution on cosmic timescales to their unique role in helping us
determine the shape of the stellar initial mass function and
validating theories of stellar evolution. Observational, theoretical
and numerical studies have made rapid progress to date, yet most
studies focus only on the properties of the populations of single
stars, while binary systems remain largely overlooked.

This is perhaps understandable from a practical perspective:
characterizing binaries in distant star clusters is difficult and
including them in numerical simulations carries a significant
computational cost, to the extent that even the most advanced $N$-body
runs often end in unstable configurations and thus lead to crashed
programs. In addition, from a theoretical perspective, it is largely
unclear how binary systems affect a cluster's color--magnitude diagram
(CMD) once we divert our attention from the fairly well-understood
main-sequence--main-sequence (MS--MS), MS--red-giant-branch (RGB), and
RGB--RGB binaries, i.e., the ubiquitous `normal,' detached
binaries. In the latter context, the formation scenarios of so-called
`blue straggler' systems are now finally reaching a concensus-like
state, with most scientists agreeing that these objects have likely
originated from either binary mergers or mass transfer between
components in binary systems.

At the same time, studies of both the field stellar population in the
solar nighborhood and the population properties of star-forming
regions and some open clusters imply that the binary fractions at the
time of starbirth are non-negligible. Mason et al. (1998) derived a
binary fraction of $\sim$70\% in massive field stars, although the
exact numbers depend on the stellar spectral type considered (Raghavan
et al. 2010). In a clustered context, Sana \& Evans (2011) and Sana et
al. (2011, 2012a,b) found a binary fraction of around 50\% in young
open clusters, including in the 30 Doradus region, which ties in very
well with the $\sim$55\% binary fraction quoted by Kobulnicky et
al. (2014) for the Cygnus OB2 association. Kouwenhoven et al. (2005)
studied the Scorpius OB2 association in detail and provided similar
binary fractions, notably separated as a function of stellar spectral
type. Binary systems are thus apparently ubiquitous in young
star-forming regions, including in regions of clustered star
formation, and hence one can expect them to have noticeable effects on
the dynamics of their host stellar systems.

\section{Characterizing binary systems in distant star clusters}

Characterizing binary systems in crowded cluster environments is
challenging, particularly in distant star clusters. Nevertheless,
careful statistical analysis of high-quality CMDs enables quantitative
characterization of binary fractions in clusters at distances out to
the Magellanic Clouds, 50--60 kpc (e.g., de Grijs et al. 2014; de
Grijs \& Bono 2015), using high-resolution {\sl Hubble Space
  Telescope} imaging. Unresolved MS--MS binaries captured in a single
resolution element will be more luminous than single stars of the same
type as the binary's primary component. In fact, unresolved binary
systems will attain brightnesses that vary between those defined by
the single-star MS and the offset sequence defined by equal-mass
(equal-luminosity) MS--MS binary systems. The latter are defined by a
binary mass ratio, $q \equiv m_2/m_1$, of unity, where $m_1$ and $m_2$
are the binary's primary and secondary components, respectively, and
$m_1 \ge m_2$. Equal-mass binaries attain luminosities that are twice
as bright as those of the equivalent single stars, which thus
translates in a binary sequence in a cluster CMD that is offset by
0.75 mag toward brighter magnitudes.\footnote{Note that while the
  offset in magnitude is most important, for binaries with $q < 1$ a
  small offset in color is also implied. The extent of the latter
  depends on both the actual mass ratio (e.g., Elson et al. 1998,
  their Fig. 3) and the photometric color examined.} In other words,
the presence of a significant fraction of binary systems will broaden
a cluster's MS ridge line. If we can accurately characterize that
broadening, we will be able to derive the most likely binary fraction.

In practice, characterization of the MS broadening is best done using
Monte Carlo simulations (e.g., Hu et al. 2010, their Fig. 6; Milone et
al. 2010, their Fig. 4; and references therein), while also allowing
for the unavoidable effects owing to photometric uncertainties and
line-of-sight blending. Given that the most important observational
effect of the presence of unresolved binary systems in a star
cluster's CMD is a spread in the magnitude direction, characterization
of the MS broadening is most easily accomplished where the MS is
shallowest -- and for young clusters, that shallowing of the MS occurs
well below the MS turn-off region, so that the effects of rapid
stellar rotation do not (yet) dominate.

Hu et al. (2010) were among the first to apply this technique to young
massive clusters. For their object of interest, the $\sim 2.8 \times
10^4 M_\odot$ compact, $\sim 15$--30 Myr-old cluster NGC 1818 in the
Large Magellanic Cloud (LMC), they derived a best-fitting MS (F-star)
binary fraction of 55\% $\pm 10$\% for $q \ge 0.4$. Depending on the
mass-ratio distribution governing these binary systems, the cluster's
overall binary fraction could be as high as 100\% (Hu et
al. 2010). This study thus established that young star clusters may
indeed host large numbers of binary systems. These binaries are
expected to result in a significant dynamical signature, thus implying
that analysis of $N$-body simulations of NGC 1818-like clusters
without any implementation of binary systems would lead to unreliable
physical insights.

\section{Radial migration and binary dissolution}

While Hu et al. (2010) obtained a first, firm constraint on a young
star cluster's close-to-primordial binary fraction, from a dynamical
perspective determination of the distribution and impact of these
systems as a function of clustercentric radius would be more
valuable. Intuitively, one might expect binary systems to behave as
single objects in an $N$-body sense, with a somewhat larger cross
section than the equivalent single stars. Realistic initial
conditions, e.g., a fractal-like stellar distribution determined by
the turbulent structure of the interstellar medium (e.g., Goodwin \&
Whitworth 2004; Goodwin et al. 2004a,b), were found to lead to rapid
dynamical redistribution of the initial stellar masses -- on
timescales of $\lesssim$ a few Myr (e.g., Allison et al. 2009, 2010)
-- and hence to a mass-segregated configuration in the inner cluster
regions.

Numerical simulations based on realistic initial conditions (i.e.,
initial substructure and dynamically cool initial configurations)
suggest that dynamical mass segregation, at least of the most massive
stars, is likely to happen in a crossing time, which is equivalent to
the free-fall time defined by the cluster's gravitational
potential. Given that NGC 1818 is 15--30 Myr old (or $\sim$5--30
crossing times), one would thus naively expect the cluster to be
(dynamically) mass segregated in its core (for confirmation, see de
Grijs et al. 2002a,b), with an increasing fraction of binary systems
at smaller radii.

However, the F-type stars studied by Hu et al. (2010), with masses of
1.3--1.6 $M_\odot$, are not expected to have already reached a state
close to energy equipartition across the cluster, because for these
stellar masses the half-mass relaxation time is much longer than the
cluster's age. This process of clusterwide dynamical mass segregation
is therefore likely still fully underway. However, contrary to
dynamical expectations based on initial conditions adopting Plummer
spheres, in de Grijs et al. (2013) we first found a hint of an
increasing fraction of binary systems in NGC 1818 from the innermost
(core) radius out to approximately its half-light radius, which we
subsequently confirmed in Li et al. (2013b). In light of our naive
initial insights, this was surprising. In fact, a previous,
preliminary study of the binary population in NGC 1818 (Elson et
al. 1998) had reached the opposite conclusion for binaries with
primary masses between $2 M_\odot$ and $5.5 M_\odot$ and $q \gtrsim
0.7$. From the core to a clustercentric distance equivalent to about 3
core radii, they found a decrease in the binary fraction from about
35\% to 20\%. We attributed this discrepant result to blending and the
near-vertical morphology of the stellar main sequence within the
magnitude range observed by Elson et al. (1998). In Geller et
al. (2015; their Fig. 7) we showed that these results are not mutually
exclusive in view of the different stellar mass ranges examined by
Elson et al. (1998) and Li et al. (2013b).

In Li et al. (2013b), we used a combination of isochrone fitting and
statistical $\chi^2$ minimization to re-investigate the binary
fraction as a function of radius in NGC 1818, in essence improving the
method applied by de Grijs et al. (2013). We also applied the same
method to NGC 1805, a similarly aged compact LMC cluster. Our results
in Li et al. (2013b) exhibit opposite trends as to the binary fractions
as a function of radius in the inner regions of both clusters. The
binary fraction in NGC 1805 decreases significantly from the cluster's
inner core to its periphery. We concluded that while early dynamical
mass segregation and the disruption of soft binary systems should be
at work in both clusters, time-scale arguments imply that early
dynamical mass segregation should be very efficient and, hence, likely
dominates the dynamical processes in the core of NGC 1805. Meanwhile,
in NGC 1818, the behavior in the core is probably dominated by
disruption of `soft' binary systems, i.e., generally wide binary
systems characterized by high kinetic compared to potential
energies. We speculate that this may be owing to the higher velocity
dispersion in the NGC 1818 core, which creates an environment in which
the efficiency of binary disruption is high compared with that of the
NGC 1805 core. In other words, by applying the same technique to two
young massive clusters in the LMC, we found two opposite radial trends
in their binary fractions. This implies that these trends are not
driven by the analysis method employed (i.e., our approach is unlikely
to introduce a bias one way or the other), but that these trends
represent physical reality.

\section{Physical implications}

Our speculation that the puzzling radial trend in NGC 1818 may have
been caused by the disription of soft binaries systems was
subsequently confirmed on the basis of $N$-body simulations (Geller et
al. 2013). Using a grid of sophisticated $N$-body simulations, in
Geller et al. (2013) we showed that the observed surface density
profile of NGC 1818 (Mackey \& Gilmore 2003) and the radial dependence
of the binary frequency (de Grijs et al. 2013; Li et al. 2013b) can be
reproduced simultaneously at the cluster's most likely age using
$N$-body simulations with both initially smooth or substructured and
equilibrium or collapsing stellar populations. The radial distribution
of the binary frequency in a rich star cluster can transition smoothly
over time from a uniform primordial radial distribution, to one that
decreases toward the core at early ages, to one that rises toward the
core at later ages. Thus both rising and falling radial distributions
in binary frequency can arise naturally from the evolution of a binary
population within the same rich star cluster as a consequence of both
dynamical disruption and mass segregation of the binaries. Indeed, we
managed to reproduce the opposite radial trend observed for NGC 1805
in a subsequent $N$-body study (Geller et al. 2015), drawing from the
same initial binary distributions in masses and orbital parameters,
although with a different initial cluster mass and half-mass radius.

Our recent analysis of the binary fractions in the young massive LMC
clusters NGC 1805 and NGC 1818, as a function of radius, thus confirms
the dynamical importance of binary systems in such dense environments
(see also Schneider et al. 2014). This thus elevates previously held
convictions that binary dynamics are predominantly important in the
context of blue straggler formation to a whole new level. We strongly
encourage the community, in particular our colleagues with expertise
in $N$-body simulations, to take up the gauntlet and henceforth fully
embrace the opportunities afforded by inclusion of binary systems in
their simulations.

\section{Binaries as diagnostic tool to trace a cluster's dynamical state}

Finally, in Geller et al. (2013), we stumbled upon an interesting
diagnostic that could potentially indicate a cluster's dynamical
state: see their Fig. 2. Our simulations started from a flat
distribution of our  NGC 1818-like model binary fraction as a function
of radius. We found that after one crossing time, the binary fraction
decreases toward the cluster core because of the disruption of wide
binaries. The higher velocity dispersion in the cluster core compared
with its periphery causes binaries in the core to move more rapidly,
on average, relative to other stars than their counterparts at larger
radii. Therefore, encounters in the core are more energetic and can
disrupt tighter binaries, while the higher density results in a higher
encounter rate in the core (Leigh \& Sills 2011).

Over time, cluster-wide mass segregation begins to dominate the radial
dependence of the binary fraction. As a consequence, the theoretical
`$r_{\rm min}$' values or `zones of avoidance' (e.g., Mapelli et
al. 2004; Ferraro et al. 2012) move radially outward. These radii
represent the clustercentric distances inside of which the local
dynamical friction timescale (Binney \& Tremaine 1987) for a binary
with a mass equivalent to that of the mean binary mass in the cluster
is shorter than the simulated time. In other words, $r_{\rm min}$
predicts the radius inside of which the binaries should experience the
effects of dynamical friction and, therefore, migrate toward the
cluster center. As a consequence, the binary fraction in the core
increases at the expense of that in the cluster's periphery. The
observed value of $r_{\rm min}$ is thus a potential indicator of the
degree and extent of mass segregation experienced by a given star
cluster.

This behavior of the radial distribution of a cluster's binary
fraction is reminiscent of that seen for the radial distribution of
blue straggler stars in many Milky Way globular clusters (e.g.,
Ferraro et al. 2012) as well as in old massive LMC clusters (Li et
al. 2013a). Ferraro et al. (2012) indeed used the minima in the radial
blue straggler frequency as an indication of the host cluster's
dynamical state. They also pointed out that blue straggler formation
in a cluster's core may have proceeded differently from that in its
outer regions. In view of the higher density in the core and the
timescales involved in blue straggler formation, core blue stragglers
most likely formed through direct stellar collisions, perhaps mediated
by binary encounters, while their peripheral counterparts may well
have formed less violently and more slowly through mass transfer
between both components of their progenitor binary systems. Once
again, this underscores the dynamical importance of binary systems on
cluster-wide scales, thus reinforcing our message to take the
contributions of binary systems seriously in the context of star
cluster dynamics.

\acknowledgements 

RdG and CL acknowledge financial support from the National Natural
Science Foundation of China (grants 11073001 and 11373010); RdG also
acknowledges travel support from the IAU to attend this conference. CL
is supported by Strategic Priority Research Program `The Emergence of
Cosmological Structures' of the Chinese Academy of Sciences (grant
XDB09000000). AMG is funded by a National Science Foundation Astronomy
and Astrophysics Postdoctoral Fellowship under Award
No.\ AST-1302765. RdG wishes to thank the organizers of IAU Symposium
316 for inviting him to give a review talk, of which this paper is a
summary. He specifically thanks Corinne Charbonnel for her competent
organizational leadership of the Scientific Organizing Committee.


\begin{thebibliography}{}

\bibitem[Allison \etal\ (2009)]{2009MNRAS.395.1449A} Allison, R.~J.,
  Goodwin, S.~P., Parker, R.~J., et al.\ 2009, {\it MNRAS}, 395, 1449

\bibitem[Allison \etal\ (2010)]{2010MNRAS.407.1098A} Allison, R.~J.,
  Goodwin, S.~P., Parker, R.~J., Portegies Zwart, S.~F., \& de Grijs,
  R.\ 2010, {\it MNRAS}, 407, 1098

\bibitem[Binney \& Tremaine(1987)]{BT87} Binney, J., \& Tremaine,
  S. 1987, {\it Galactic Dynamics} (Princeton, NJ: Princeton
  Univ. Press)

\bibitem[de Grijs \etal\ (2002a)]{2002MNRAS.331..228D} de Grijs, R.,
  Johnson, R.~A., Gilmore, G.~F., \& Frayn, C.~M.\ 2002a, {\it MNRAS},
  331, 228

\bibitem[de Grijs \etal\ (2002b)]{2002MNRAS.331..245D} de Grijs, R.,
  Gilmore, G.~F., Johnson, R.~A., \& Mackey, A.~D.\ 2002b, {\it
    MNRAS}, 331, 245

\bibitem[de Grijs \etal\ (2013)]{2013ApJ...765....4D} de Grijs, R.,
  Li, C., Zheng, Y., et al.\ 2013, {\it ApJ}, 765, 4

\bibitem[de Grijs \etal\ (2014)]{2014AJ....147..122D} de Grijs, R.,
  Wicker, J.~E., \& Bono, G.\ 2014, {\it AJ}, 147, 122

\bibitem[de Grijs \& Bono(2015)]{2015AJ....149..179D} de Grijs, R., \&
  Bono, G.\ 2015, {\it AJ}, 149, 179

\bibitem[Elson \etal\ (1998)]{1998MNRAS.300..857E} Elson, R.~A.~W.,
  Sigurdsson, S., Davies, M., Hurley, J., \& Gilmore, G.\ 1998, {\it
    MNRAS}, 300, 857

\bibitem[Ferraro et al.(2012)]{2012Natur.492..393F} Ferraro, F.~R.,
  Lanzoni, B., Dalessandro, E., et al.\ 2012, {\it Nature}, 492, 393

\bibitem[Geller \etal\ (2013)]{2013ApJ...779...30G} Geller, A.~M., de
  Grijs, R., Li, C., \& Hurley, J.~R.\ 2013, {\it ApJ}, 779, 30

\bibitem[Geller \etal\ (2015)]{2015ApJ...805...11G} Geller, A.~M., de
  Grijs, R., Li, C., \& Hurley, J.~R.\ 2015, {\it ApJ}, 805, 11

\bibitem[Goodwin \& Whitworth(2004)]{2004A&A...413..929G} Goodwin,
  S.~P., \& Whitworth, A.~P.\ 2004, {\it A\&A}, 413, 929

\bibitem[Goodwin \etal\ (2004a)]{2004A&A...414..633G} Goodwin, S.~P.,
  Whitworth, A.~P., \& Ward-Thompson, D.\ 2004a, {\it A\&A}, 414, 633

\bibitem[Goodwin \etal\ (2004b)]{2004A&A...423..169G} Goodwin, S.~P.,
  Whitworth, A.~P., \& Ward-Thompson, D.\ 2004b, {\it A\&A}, 423, 169

\bibitem[Hu \etal\ (2010)]{2010ApJ...724..649H} Hu, Y., Deng, L., de
  Grijs, R., Liu, Q., \& Goodwin, S.~P.\ 2010, {\it ApJ}, 724, 649

\bibitem[Kobulnicky \etal\ (2014)]{2014ApJS..213...34K} Kobulnicky,
  H.~A., Kiminki, D.~C., Lundquist, M.~J., et al.\ 2014, {\it ApJS},
  213, 34

\bibitem[Kouwenhoven \etal\ (2005)]{2005A&A...430..137K} Kouwenhoven,
  M.~B.~N., Brown, A.~G.~A., Zinnecker, H., Kaper, L., \& Portegies
  Zwart, S.~F.\ 2005, {\it A\&A}, 430, 137

\bibitem[Leigh \& Sills(2011)]{2011MNRAS.410.2370L} Leigh, N., \&
  Sills, A.\ 2011, {\it MNRAS}, 410, 2370

\bibitem[Li et al.(2013a)]{2013ApJ...770L...7L} Li, C., de Grijs, R.,
  Deng, L., \& Liu, X.\ 2013a, {\it ApJL}, 770, L7

\bibitem[Li \etal\ (2013b)]{2013MNRAS.436.1497L} Li, C., de Grijs, R.,
  \& Deng, L.\ 2013b, {\it MNRAS}, 436, 1497

\bibitem[Mackey \& Gilmore(2003)]{2003MNRAS.338...85M} Mackey, A.~D.,
  \& Gilmore, G.~F.\ 2003, {\it MNRAS}, 338, 85

\bibitem[Mapelli et al.(2004)]{2004ApJ...605L..29M} Mapelli, M.,
  Sigurdsson, S., Colpi, M., et al.\ 2004, {\it ApJL}, 605, L29

\bibitem[Mason \etal\ (1998)]{1998AJ....116.2975M} Mason, B.~D.,
  Henry, T.~J., Hartkopf, W.~I., ten Brummelaar, T., \& Soderblom,
  D.~R.\ 1998, {\it AJ}, 116, 2975

\bibitem[Milone \etal\ (2010)]{2010ApJ...709.1183M} Milone, A.~P.,
  Piotto, G., King, I.~R., et al.\ 2010, {\it ApJ}, 709, 1183

\bibitem[Raghavan et al.(2010)]{2010ApJS..190....1R} Raghavan, D.,
  McAlister, H.~A., Henry, T.~J., et al.\ 2010, {\it ApJS}, 190, 1

\bibitem[Sana \& Evans(2011)]{2011IAUS..272..474S} Sana, H., \& Evans,
  C.~J.\ 2011, {\it IAU Symp.}, 272, 474

\bibitem[Sana \etal\ (2011)]{2011MNRAS.416..817S} Sana, H., James, G.,
  \& Gosset, E.\ 2011, {\it MNRAS}, 416, 817

\bibitem[Sana \etal\ (2012a)]{2012Sci...337..444S} Sana, H., de Mink,
  S.~E., de Koter, A., et al.\ 2012a, {\it Science}, 337, 444

\bibitem[Sana \etal\ (2012b)]{2012ASPC..465..284S} Sana, H., Dunstall,
  P.~R., H{\'e}nault-Brunet, V., et al.\ 2012b, in: {\it
    Proc. Scient. Mtg in Honor of Anthony F.~J.~Moffat}, ASP
  Conf. Ser. 465, 284

\bibitem[Schneider \etal\ (2014)]{2014ApJ...780..117S} Schneider,
  F.~R.~N., Izzard, R.~G., de Mink, S.~E., et al.\ 2014, {\it ApJ},
  780, 117

\end{thebibliography}
\end{document}